\documentclass[
    aps,
    superscriptaddress,
    tightenlines,
    nofootinbib,
    floatfix,
    longbibliography,
    notitlepage,
    twocolumn,
    preprintnumbers
]{revtex4-1}

\usepackage{geometry}
\usepackage[T1]{fontenc}
\usepackage[utf8]{inputenc}
\usepackage[UKenglish]{babel}
\usepackage{siunitx}
\usepackage{physics}
\usepackage[dvipsnames]{xcolor}
\usepackage{listings}
\usepackage[super]{nth}
\usepackage{mathtools}
\usepackage{nicefrac}
\usepackage{amsmath}
\usepackage{amssymb}
\usepackage{amsfonts}
\usepackage{dsfont}
\usepackage{upgreek}
\usepackage{hyperref}
\usepackage{booktabs}
\usepackage{comment}

\usepackage[caption=false,position=top,labelformat=parens]{subfig}

\usepackage{tikz}
    \usetikzlibrary{
        mindmap,
        shadows,
        arrows,
        matrix,
        arrows.meta,
        shapes,
        decorations,
        decorations.shapes,
        decorations.markings,
        decorations.pathmorphing
    }
    \tikzstyle{arrow} = [very thick,->,>=stealth]
    \tikzstyle{middleArrow} = [-, decoration={markings, mark=at position 0.5 with {\arrow{Stealth}} }, line width = 0.1em, postaction={decorate} ]
    \tikzstyle{snakeLine} = [-, decorate, decoration={snake,amplitude=0.2em,segment length=0.6em}, line width = 0.1em, ]
    \tikzstyle{latticeLine} = [-, line width=0.1em]
    \tikzstyle{latticeLink} = [-, decoration={markings, mark=at position 0.6 with {\arrow{Stealth}} }, line width = 0.1em, fzjred, postaction={decorate} ]

\usepackage{ulem}
\usepackage{placeins} 
\newcommand{\im}{\ensuremath{\mathrm{i}}}

\newcommand{\DLR}{\ensuremath{\overset{\leftrightarrow}{D}}}

\renewcommand{\O}{\ensuremath{\mathcal{O}}}

\newcommand{\Ncfg}{\ensuremath{N_{\mathrm{cfg}}}}

\renewcommand{\bar}[1]{\ensuremath{\overline{#1}}}

\newcommand{\COpt}[3][]{\ensuremath{\mathrm{C}_{3\mathrm{pt}}^{#1}\left(#2,\,#3\right)}}

\newcommand{\Cpt}[1]{\ensuremath{}\mathrm{C}_{2\mathrm{pt}}\left(#1\right)}

\definecolor{fzjblue}{RGB}{2,61,107}
\definecolor{fzjlightblue}{RGB}{173,189,227}
\definecolor{fzjgray}{RGB}{235,235,235}
\definecolor{fzjred}{RGB}{235, 95, 115}
\definecolor{fzjgreen}{RGB}{185, 210, 95} 
\definecolor{fzjyellow}{RGB}{250, 235, 90}
\definecolor{fzjviolet}{RGB}{175, 130, 185}
\definecolor{fzjorange}{RGB}{250, 180, 90}
\definecolor{fzjblack}{RGB}{0,0,0}
\definecolor{fzjwhite}{RGB}{255,255,255}

\newcommand{\jsc}{
    JARA \& J\"{u}lich Supercomputing Center,
    Forschungszentrum J\"{u}lich, 54245 J\"{u}lich, Germany
}
\newcommand{\ias}{
    Institute for Advanced Simulation,
    Forschungszentrum J\"{u}lich, 54245 J\"{u}lich, Germany
}
\newcommand{\casa}{
    Center for Advanced Simulation and Analytics (CASA),
    Forschungszentrum Jülich, 52425 J\"{u}lich, Germany
}
\newcommand{\bonn}{
    Helmholtz-Institut f\"{u}r Strahlen- und Kernphysik,
    Rheinische Friedrich-Wilhelms-Universit\"{a}t Bonn, 53115 Bonn, Germany
}
\newcommand{\mitctp}{
    Center for Theoretical Physics,  
    MIT,  Cambridge, MA, USA
}
\newcommand{\suny}{
    Department of Physics and Astronomy, 
    Stony Brook University, Stony Brook, NY, USA
}
\newcommand{\bnl}{
    RIKEN/BNL  Research  Center,  
    Brookhaven  National  Laboratory,  Upton, NY, USA
}
\newcommand{\nmsu}{
    Department of Physics,
    New Mexico State University, Las Cruces, NM, USA
}
\newcommand{\UA}{
    Department of Physics,
    University  of  Arizona,  Tucson,  AZ, USA
}
\newcommand{\zppt}{
    Deutsches Elektronen-Synchrotron DESY,
    Platanenallee 6, 15738 Zeuthen, Germany
}

\date{\today}

\begin{document}

\preprint{DESY-23-072}

\title{Moments of Parton Distributions Functions from Lattice QCD at the Physical Point}
\author{Marcel~Rodekamp}
   \affiliation{\jsc}
   \affiliation{\ias}
   \affiliation{\bonn}
   \affiliation{\casa}
\author{Michael~Engelhardt}
   \affiliation{\nmsu}
\author{Jeremy~R.~Green}
   \affiliation{\zppt}
\author{Stefan~Krieg}
   \affiliation{\jsc}
   \affiliation{\ias}
   \affiliation{\bonn}
   \affiliation{\casa}
\author{Stefan~Meinel}
   \affiliation{\UA}
\author{John~W.~Negele}
   \affiliation{\mitctp}
\author{Andrew~Pochinsky}
    \affiliation{\mitctp}
\author{Sergey~Syritsyn}
    \affiliation{\suny}
    \affiliation{\bnl}
\begin{abstract}
We present a Lattice QCD calculation of the second Mellin moments of the nucleon axial, vector and tensor parton distribution functions (PDFs).
The calculation is performed at the physical pion mass with two different lattice spacings, and includes both zero and non-zero nucleon momenta.
In our preliminary analysis, we identify operators that greatly reduce excited-state contamination.
\par\bigskip\noindent
DIS2023: XXX International Workshop on Deep-Inelastic Scattering and Related Subjects, \\
Michigan State University, USA, 27-31 March 2023 \\
\begin{center}
\includegraphics[width=9cm]{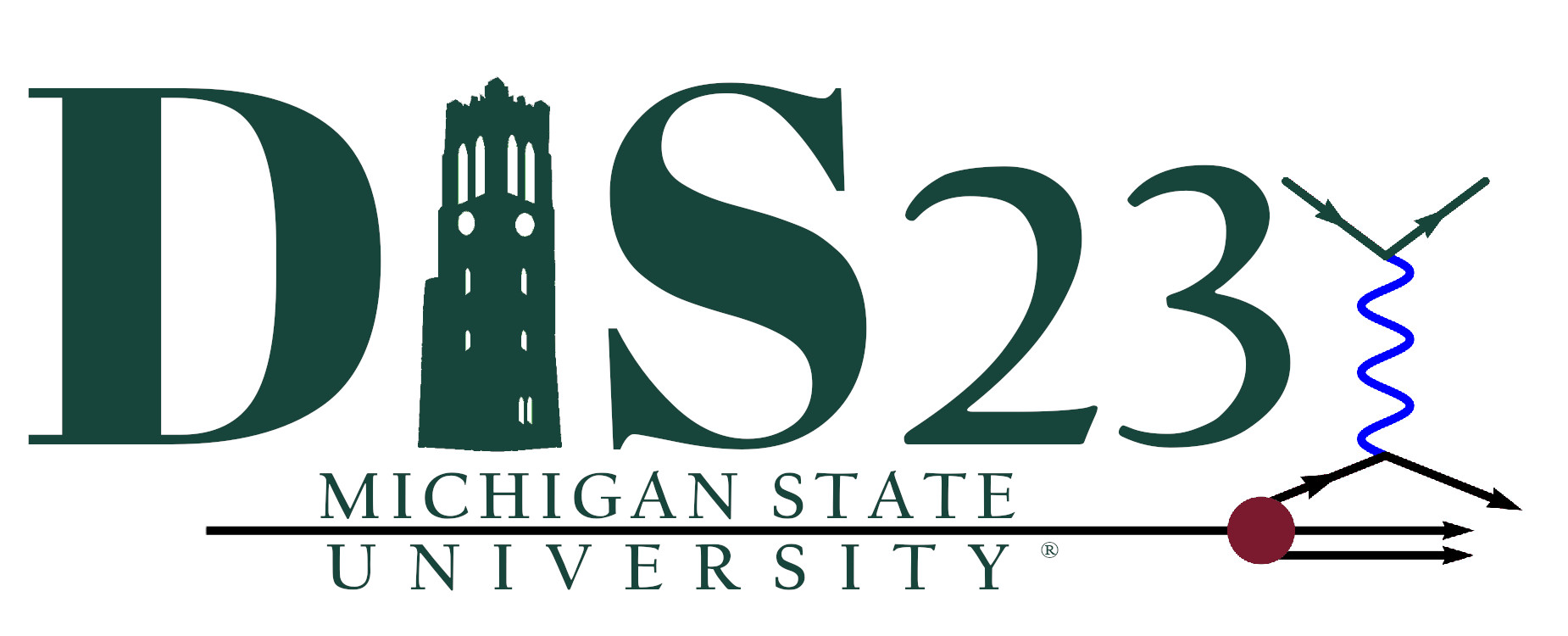}
\end{center}
\end{abstract}
 
\maketitle

\section{Introduction}\label{sec-Introduction}
Parton distribution functions (PDFs) have proved to be a valuable tool 
in describing the structure of hadrons and making predictions for high-energy processes at hadron colliders.
First-principles calculations of PDFs are very difficult due to their non-perturbative nature.
Lattice QCD provides a way of calculating (non-perturbative) observables 
by introducing a four-dimensional Euclidean hypercubic lattice to discretise the space-time, serving as a regulator.
The path integral is then calculated with a Monte Carlo algorithm.

In the past years the Lattice QCD community has made tremendous progress 
in calculating PDFs by directly assessing their Bjorken-$x$ dependence 
from the leading-twist contribution to bilocal matrix elements at high momentum.
In this work, we concentrate on 
the second Mellin moment $\expval{x}$~\cite{haglerHadronStructureLattice2010,linPartonDistributionsLattice2018,constantinouXdependenceHadronicParton2021}
via matrix elements of local twist-two operators,
which does not require large momenta to suppress higher-twist contributions
and thus simplifies the numerical estimation. 
We aim to understand the excited-state contamination and identify a set of matrix elements that have particularly low contributions from exited states.
This requires the study of matrix elements at finite but small momenta as some have contributions 
only at non-zero momentum.
The study of forward matrix elements of local operators at non-zero momentum 
is uncommon but has been done in references~\cite{brattNucleonStructureMixed2010,therqcdcollaborationNucleonAxialStructure2020,barcaPiMatrixElements2022}.

This contribution is organized as follows. 
In section \ref{sec-Method} we explain our analysis chain and discuss in detail 
which operators are considered.
In section \ref{sec-Results} we show our preliminary results of the different steps of the 
analysis and discuss their significance in terms of excited-state contamination.
Last, in \ref{sec-Summary} we summarize our findings.

\section{Method}\label{sec-Method}
Moments of PDFs can be obtained by calculating forward matrix elements of 
local leading twist operators \cite{martinelli1988lattice,martinelli1989lattice,gockelerLatticeOperatorsMoments1996,alexandrou2016parton}
\begin{equation}
    \O^X \equiv \O^X_{\{ \alpha,\mu\}} =  \bar{q} \Gamma^X_{\{ \alpha} \DLR^{\phantom{X}}_{\mu \}} q, \label{eq-leading-twist-operator}
\end{equation}
where $X = V,A,T$ indicates the vector, axial or tensor channel leading to unpolarized,
polarized or transversity PDFs respectively.
We symmetrize the indices and take the traceless part, denoted by $\{\alpha,\mu\}$,
and restrict ourselves to the isovector channel, $\O^X(q=u)-\O^X(q=d)$, to avoid calculating disconnected diagrams.
The left-right acting covariant derivative $\DLR$ is constructed on the Euclidean lattice by 
finite differences of neighbouring points connected by an appropriate gauge link $U_{\mu}(\mathbf{x})$.
One can show that the forward matrix element is proportional to the desired moment $\expval{x}$~\cite{gockelerQuarkHelicityFlip2005,haglerHadronStructureLattice2010,alexandrou2016parton}
\begin{equation}
\begin{aligned}
    \mathcal{M} &\equiv \mel{N(p)}{\O^X_{\{\alpha,\mu\}}}{N(p)} \\ 
                &= \expval{x} \bar{u}_{N(p)} \Gamma^X_{\{ \alpha} \im p^{\phantom{X}}_{\mu \}} u_{N(p)},\label{eq-matrix-element}
\end{aligned}
\end{equation}
where $p$ is the nucleon's 4-momentum. 

In the continuum, the operators~\eqref{eq-leading-twist-operator} are classified 
according to irreducible representations of the Lorentz group, which in Euclidean space
is replaced by the orthogonal group~\cite{gockelerLatticeOperatorsMoments1996}.
On the lattice, the orthogonal group is further reduced to the hypercubic group $H(4)$.
This explicit breaking causes some operators to mix with lower-dimensional ones; 
however, for a one-derivative operator as used here this does not happen. 
Still, the Euclidean irreducible representations to which our operators belong 
split into multiple hypercubic irreps; 
we use the typical notation where $\tau_a^{(b)}$ denotes the $a$th $b$-dimensional irrep.
Each of the latter has a different renormalization factor, so we construct operators with definite hypercubic irreducible representation to keep the renormalization diagonal~\cite{gockelerLatticeOperatorsMoments1996}.
In practice this means for each $\tau_{a}^{(b)}$ 
we have to calculate the renomalization factor $Z_{\tau_{a}^{(b)}}$ to multiply
matrix elements of an operator that transforms irreducibly under it, 
consequently we denote $Z_{\O^X} \equiv Z_{\tau_{a}^{(b)}}$.

\begin{figure}
\resizebox{\linewidth}{!}{
\begin{tikzpicture}
    \node[fill=fzjviolet, draw=fzjviolet, circle, font=\large ] at (0,0)  (NL) {N(P)};
    \node[fill=fzjviolet, draw=fzjviolet, circle, font=\large ] at (10,0) (NR)  {N(P)};

    \draw[fzjred  , line width = 2pt, line cap = round] (NL.north east) to[out=25,in=155] (NR.north west);
    \draw[fzjgreen, line width = 2pt, line cap = round] (NL.south east) to[out=335,in=205] (NR.south west);
    \draw[fzjblue , line width = 2pt, line cap = round] (NL.east) -- (NR.west);

    \node[fill=fzjlightblue, draw=fzjblue, ellipse, minimum height = 4em, font=\large] at (5,1.2) {$\O^X$};

    \draw[arrow] (-1,-2) -- (11,-2);
    \node[font=\large] at (0, -2.3) {0};
    \node[font=\large] at (5, -2.3) {$\tau$};
    \node[font=\large] at (10,-2.3) {$T$};

    \node[fill=fzjviolet, draw=fzjviolet, circle, font=\large ] at (0,0)  (NL) {N(P)};
    \node[fill=fzjviolet, draw=fzjviolet, circle, font=\large ] at (10,0) (NR)  {N(P)};
\end{tikzpicture}
}
\caption{
    Graphical representation of $\COpt[\O^X]{T}{\tau}$, a source 
    nucleon inserted at time $t=0$ and a sink nucleon removed at time $t=T$.
    A local leading twist operator~\eqref{eq-leading-twist-operator} is 
    inserted on a given time slice $\tau$.
    The nucleons on the lattice are represented by interpolating operators 
    while $\O^X$ is determined by finite differences connected with gauge links. 
}\label{fig-correlation}
\end{figure}
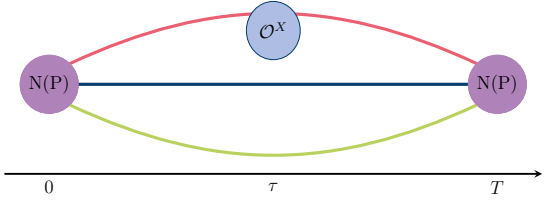

The matrix element of~\eqref{eq-matrix-element} can be obtained from the 
lattice by considering ratios of three-point and two-point correlation functions~\cite{gockelerQuarkHelicityFlip2005,alexandrou2016parton}.
The two-point correlation function $\Cpt{\tau}$ measures the correlation of a nucleon source and 
a nucleon sink separated by a time $\tau$,
while the three-point correlation function $\COpt[\O^X]{T}{\tau}$ separates the source and sink
nucleons by a time $T$ and inserts an operator of interest, here $\O^X$, at time 
$\tau$.
For a graphical representation consider figure~\ref{fig-correlation}. 
The matrix element is then obtained in the limit
\begin{align}
    \mathcal{M}
    &= \lim_{T-\tau,\tau\to\infty} R(T,\tau) \\
    &\equiv \lim_{T-\tau,\tau\to\infty} \frac{ \COpt[\O^X]{T}{\tau}}{\Cpt{T}}.
    \label{eq-ratio-definition}
\end{align}
Doing a spectral analysis of this ratio reveals the matrix element of the ground state
\begin{equation}
    R(T,\tau) = \mathcal{M} + \text{Excited States}.\label{eq-1-state}
\end{equation}
Expanding further, including the first excited state,
shows the dominant excited-state contamination
\begin{gather}
    \mathcal{M} 
    \frac{1 + c_1 e^{-\frac{T}{2}\Delta E } \cosh\left[ \left(\nicefrac{T}{2} - \tau \right) \Delta E\right] + c_2 e^{- T \Delta E} }{1 + c_3 e^{- T \Delta E}},\label{eq-2-state} 
\end{gather}
where we use $\Delta E = E_1-E_0$.
Naturally, one would consider large $T,\tau$ approaching the limit 
of~\eqref{eq-ratio-definition}. 
The statistical noise increases with $T$ , implying increased numerical costs for this approach.
The constants $c_i$ depend on the operator $\O^X$; thus, if they appear 
to be small or obey some symmetry, the excited-state contamination of the matrix element is further reduced.

Considering the sum of ratios 
\begin{equation}
\begin{gathered}
    S(T,\tau_\mathrm{skip}) = a\sum_{\tau = \tau_\mathrm{skip}}^{T-\tau_\mathrm{skip}} R(T,\tau) \\
    = \mathcal{M}\left( T-\tau_\mathrm{skip} \right) + c + \text{Excited States},
    \label{eq-definition-sum-ratios}
\end{gathered}
\end{equation}
excited-state contamination is exponentially suppressed with $T$ compared to $\nicefrac{T}{2}$ for the ratios themselves~\cite{capitani2010systematic,bulava2010b}.
Increasing $\tau_\mathrm{skip}$ reduces excited-state contamination, here typically a value of $\nicefrac{\tau_\mathrm{skip}}{a} = 1,2,3$ is enough.
In order to extract the matrix element from the ratio sums, up to excited states, we can either fit the slope according to \eqref{eq-definition-sum-ratios} or use finite differences
\begin{equation}
    \mathcal{M} = \frac{S(T+\delta, \tau_\mathrm{skip}) - S(T, \tau_\mathrm{skip})}{\delta}. 
    \label{eq-definition-FD}
\end{equation}
Due to the available data we use a combination of $\nicefrac{\delta}{a} \in\{1,2,3\}$ depending on whether a neighbour $T+\delta$ is available.

Having the basic quantities of interest, we can summarize the analysis as follows.
First estimate the ratios $R(T,\tau)$ and ratio sums $S(T,\tau_\mathrm{skip})$.
Currently, we extract matrix elements $\mathcal{M}$ in two ways.
First, fitting the slope of $S(T,\tau_\mathrm{skip})$ at fixed $\tau_\mathrm{skip}$ limiting $T\geq T'$ for various minimal source-sink separations $T'$.
Second, extracting the slope via finite differences at a source-sink separation $T=T'$.
A matrix element extracted with either those is denoted by $\mathcal{M}\vert_{T',\mathfrak{m}}$ where $\mathfrak{m}$ denotes one of the two above methods.
For both methods, as we increase $T'$ excited states are expected to decay.
Dividing the kinematic factor results in a $T'$-dependent moment for a given operator $\O^X$ and momentum $p$ using the matrix element extraction method $\mathfrak{m}$ 
\begin{equation}
    \mathfrak{X}_{\O^X,p,\mathfrak{m}}(T') = \frac{\mathcal{M}\vert_{T',\mathfrak{m}}}{\bar{u}_{N(p)} \Gamma^X_{\{ \alpha} \im p^{\phantom{X}}_{\mu \}} u_{N(p)}}.\label{eq-moment-step}
\end{equation}
To simplify the following equations, we define a compound index $j = \left(\O^X,p,\mathfrak{m}\right)$ that runs over
all operators and momenta with nonzero kinematic factors as well as the different methods to obtain the matrix element.
Determining the renormalization factors in RI-(S)MOM and matching it to \bar{\mathrm{MS}}(2 GeV) allows us to express the renormalized moment $\mathfrak{X}_{j}^\mathrm{ren}(T') = Z_{\O^X} \cdot \mathfrak{X}_{j}(T')$.
With these we define the central value as weighted average of the different results
\begin{equation}
    \expval{x}^\mathrm{ren} = 
            \sum_{\tiny j,T' \geq T_\mathrm{plat}^{j}} \mathfrak{W}_{j}(T') \mathfrak{X}_{j}^\mathrm{ren}(T') 
        .\label{eq-final-moment}
\end{equation}
Here $T_\mathrm{plat}^{j}$ denotes the smallest source-sink separation such that $\mathfrak{X}_{j}(T')$ agree for all $T'\geq T_\mathrm{plat}^{j}$.
The weights $\mathfrak{W}_{j}(T') \propto \nicefrac{1}{\sigma^2_{j}(T')}$ are normalised such that they sum to 1. 
The used variances are estimated via bootstrap over $\mathfrak{X}_{j}^{}(T')$ and the errors of the renormalization constants are propagated.
Last we estimate a systematic error by taking the weighted standard deviation over the different results
\begin{equation}
    \sigma_{syst}^2 =         
        \sum_{\tiny j, T'\geq T_\mathrm{plat}^j} \mathfrak{W}_{j}(T') \left[ \mathfrak{X}_{j}^\mathrm{ren}(T') - \expval{x}^\mathrm{ren} \right]^2
        \label{eq-syst-error}.
\end{equation}

\section{Results}\label{sec-Results}
We use a tree-level Symanzik-improved gauge action with 2+1 flavour tree-level improved Wilson Clover fermions 
coupling via 2-level HEX-smearing. 
The details can be found in~\cite{durrLatticeQCDPhysical2011,durrLatticeQCDPhysical2011a,hasanNucleonAxialScalar2019} and relevant parameters are summarized in table~\ref{tab-Simulation-Details}.
Two ensembles, coarse and fine, have been generated at the physical pion mass corresponding to lattice spacings of $\SI{0.1163(4)}{\femto\meter}$ and $\SI{0.0926(6)}{\femto\meter}$ respectively.
On each ensemble two-point and three-point correlation functions are calculated with source-sink 
separations ranging from $\approx\SI{0.3}{\femto\meter}$ to $\SI{1.4}{\femto\meter}$ and 
$\approx\SI{0.9}{\femto\meter}$ to $\approx\SI{1.5}{\femto\meter}$.
For each ensemble two momenta are chosen; $\vec{p} = (p_x,0,0)$ with $p_x = 0, -2[\nicefrac{2\pi}{L}]$ and $p_x = 0,-1 [\nicefrac{2\pi}{L}]$ respectively.

\begin{table*}
\begin{tabular}{l cccccccccc}
    \toprule
    Ensemble & Size & $\beta$ & $a[\SI{}{\femto\meter}]$ & $m_{\pi} [\SI{}{\MeV}]$ &  $m_\pi L$ & $\nicefrac{T}{a}$ & $p_x[\nicefrac{2\pi}{L}]$ & $\Ncfg$ \\
    \midrule
    Coarse & $\num{48}^4$ & \num{3.31} & \num{0.1163(4)} & \num{136(2)} & \num{3.9} & $3,4,5,6,7,8,10,12$ & $0,-2$ & \num{212} \\
    Fine   & $\num{64}^4$ & \num{3.5 } & \num{0.0926(6)} & \num{133(1)} & \num{4.0} & $10,13,16$          & $0,-1$ & \num{427} \\
    \bottomrule
\end{tabular}
\caption{
    Details of the used ensembles.
    The ensembles are at the physical pion mass, $m_{\pi} \approx m_{\pi}^{phys}$.
    A larger and a smaller lattice spacing, labelled as "Coarse" and "Fine" respectively, are available. 
    The ensembles are generated with a tree-level Symanzik-improved gauge action 
    with 2+1 flavour tree-level improved Wilson Clover fermions 
    coupled via 2-level HEX-smearing~\cite{durrLatticeQCDPhysical2011,durrLatticeQCDPhysical2011a,hasanNucleonAxialScalar2019}. 
    Furthermore, the available source-sink separations ($T$) and momenta ($p_x$) 
    which are used in the calculation of the ratios, equation~\eqref{eq-ratio-definition}, are displayed.
}
\label{tab-Simulation-Details}
\end{table*}

\begin{figure*}
\subfloat[\phantom{\rule{0.86\textwidth}{1pt}}\label{Ratios-1}]{\includegraphics[scale=0.27]{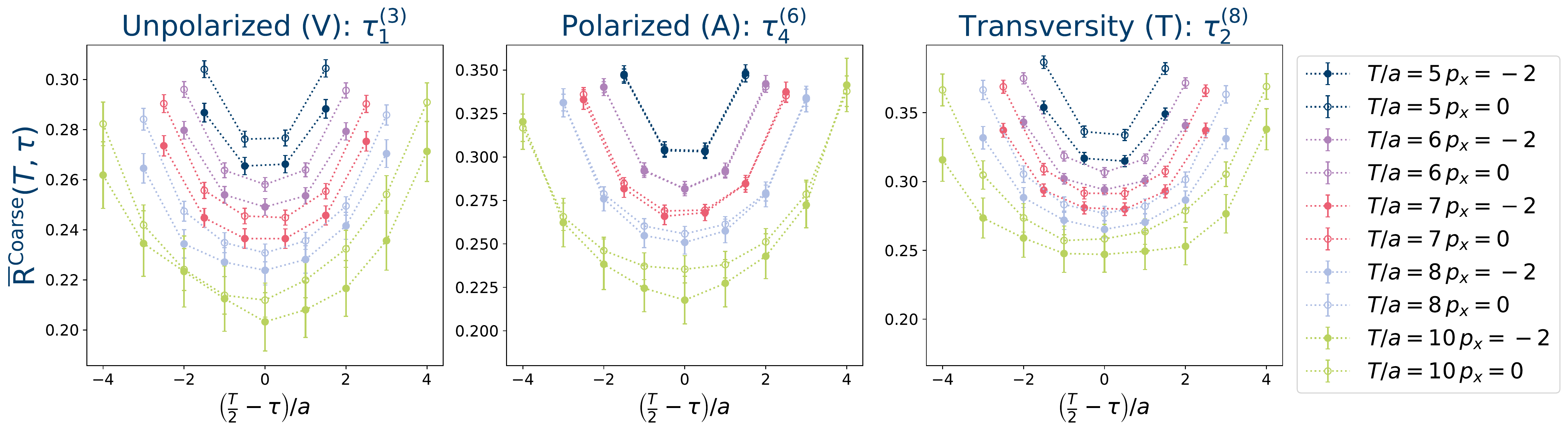}}\\
\subfloat[\phantom{\rule{0.86\textwidth}{1pt}}\label{Ratios-2}]{\includegraphics[scale=0.27]{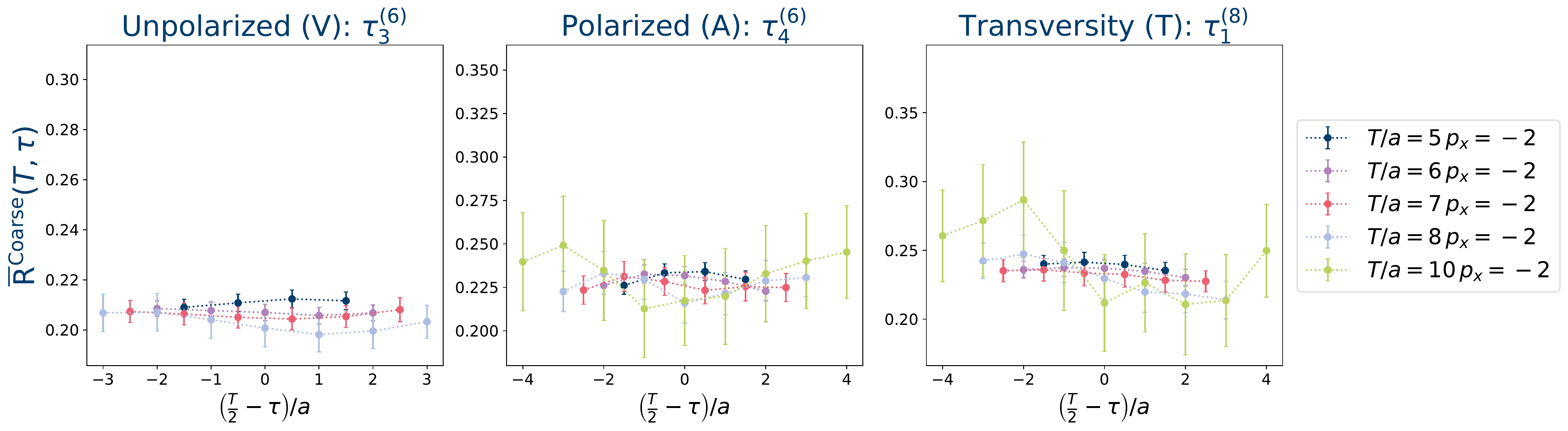}}
\caption{
Ratios, cf.\@ eq.\@~\eqref{eq-ratio-definition}, for the coarse ensemble.
Different source-sink separations $T$ are shown in different colours and the two momenta, in~\ref{Ratios-1}, are distinguished by hollow and filled markers.
Different sets of operators were chosen for the two subfigures.
}\label{Ratios}
\end{figure*}

\begin{figure*}
\includegraphics[scale=0.27]{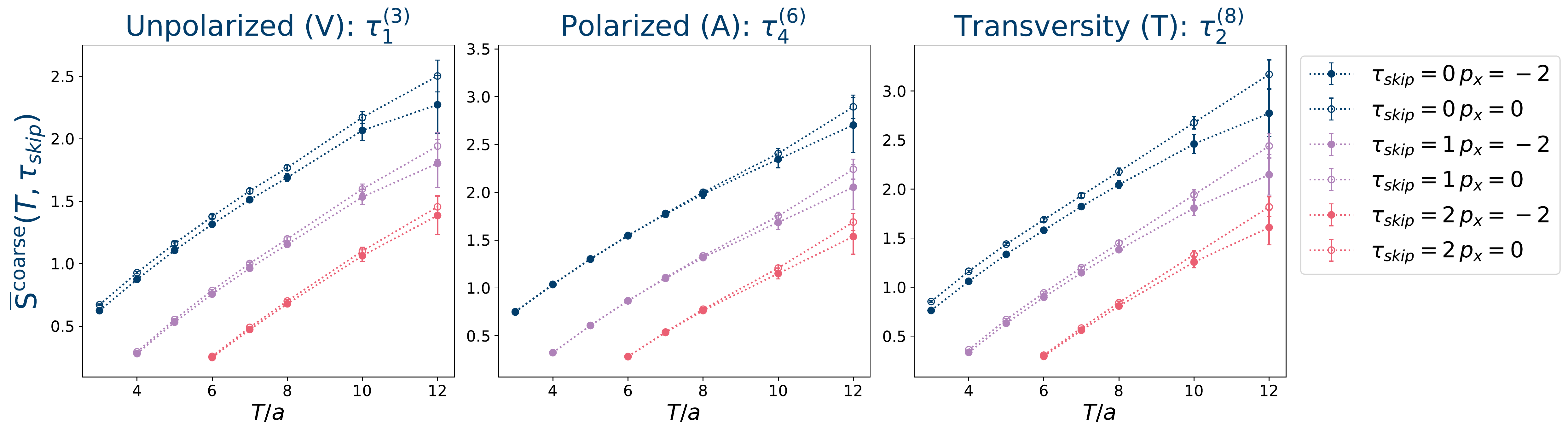}
\caption{
    ratio sums $\bar{S}(T,\tau_\mathrm{skip})$ on the coarse lattice.
    The chosen operators $\O^X$ are the same as in~\ref{Ratios-1}.
    Each $\bar{S}(T,\tau_\mathrm{skip})$ is plotted at fixed $\tau_\mathrm{skip}$, indicated by colour, over different source-sink separations.
    As in~\ref{Ratios-1} different momenta are displayed with hollow and filled markers.
}\label{Sum-Ratios}
\end{figure*}

\begin{figure*}
\resizebox{1\linewidth}{!}{
\includegraphics{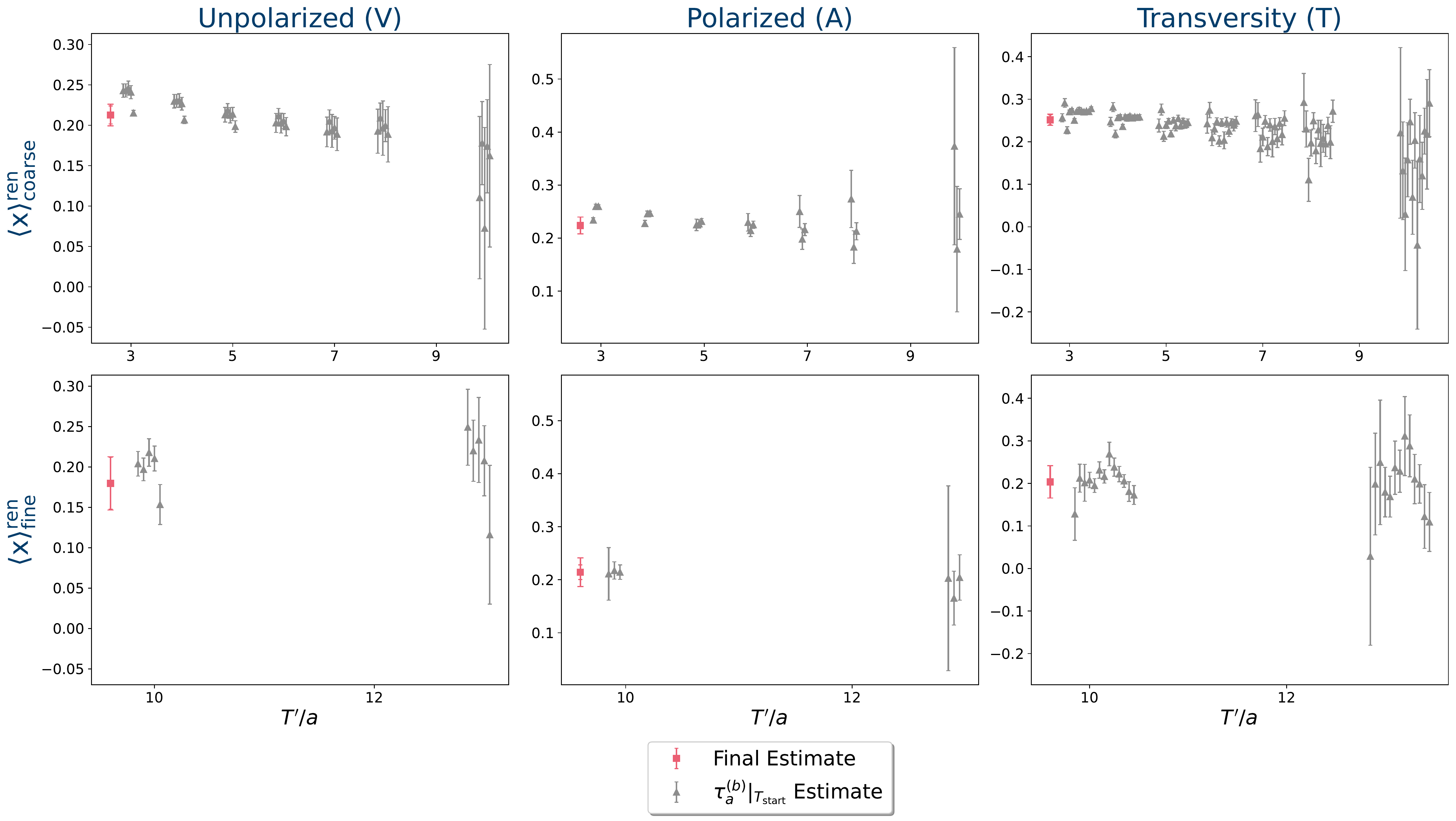}
}
\caption{
    Results for the renormalized moments computed from the ratio sums~\eqref{eq-definition-sum-ratios}.
    Moments are computed from finite differences at fixed $T'$ and $\tau_\mathrm{skip} = 1$.
    The grey points represent $\mathfrak{X}^\mathrm{ren}_{j}(T')$.
    The moments are plotted against $T'$ and slightly displaced for clarity.
    The red points represent the preliminary result obtained using~\eqref{eq-final-moment}.
    The inner and outer errorbars indicate statistical and total uncertainty respectively. 
}\label{moments}
\end{figure*}

Figures~\ref{Ratios}~to~\ref{moments} show the different steps of the analysis.
For a given channel X, the figures~\ref{Ratios-1}~and~\ref{Ratios-2} show results 
using one possible operator $\O^X_{\{\alpha,\mu\}}$.
Here we multiply with the kinematic factor $\bar{R}(T,\tau) = \nicefrac{1}{\bar{u}_{N(p)} \Gamma^X_{\{\alpha} \im p^{\phantom{X}}_{\mu\}} u_{N(p)}} \cdot R(T,\tau)$ such that a plateau corresponds to the bare moment.
These plots omit the largest source sink separation due to its enormous statistical uncertainty.
Two different operators $\O^X_{\{\alpha,\mu\}}$ are used for each channel $X$ going from figure~\ref{Ratios-1}~to~\ref{Ratios-2}.
For the axial channel ($X=A$) both operators transform under $\tau_{4}^{(6)}$.
The lower excited-state contamination for some operators can be deduced directly from these figures as~\ref{Ratios-1} obey the cosh behaviour of~\eqref{eq-2-state} while~\ref{Ratios-2} are perfectly flat within uncertainty.
Notably, those operators have a contribution only at finite momentum ($p_x \neq 0$ here) which increases the statistical noise.

A similar rescaling has been done for the ratio sums $\bar{S}(T,\tau_\mathrm{skip}) = \nicefrac{1}{\bar{u}_{N(p)} \Gamma^X_{\{\alpha}\im p^{\phantom{X}}_{\mu\}} u_{N(p)}} \cdot S(T,\tau_\mathrm{skip})$ 
in figure~\ref{Sum-Ratios}.
The exited state contamination is indicated by the slight curvature though much more obscured compared to the ratios. 
The slopes of these lines are used in the current analysis shown in figure~\ref{moments}.
In future work we want to include a 2-state analysis as in~\eqref{eq-2-state}, improving on the central value~\eqref{eq-final-moment} as well as the systematic error estimation~\eqref{eq-syst-error}. 

In Figure~\ref{moments} the gray points correspond to the different renormalized moments $\mathfrak{X}_{j}^\mathrm{ren}(T')$ from finite differences
plotted against $T'$ but slightly displaced to increase readability.
The blue points represent the preliminary result, computed using~\eqref{eq-final-moment}.
The inner errorbars represent the statistical -- bootstrap -- uncertainty while the outer ones add the estimate of systematic errors, $\sqrt{\sigma_{stat}^2 + \sigma_{syst}^2}$. 
The upper and lower row collect results from the coarse and fine ensemble respectively.
Encouragingly, the central values agree within the uncertainties.

\section{Summary}\label{sec-Summary}
We calculate the second Mellin moment $\expval{x}$ of axial, vector and tensor PDFs from lattice QCD with two lattice spacings at the physical pion mass. 
The study includes nucleon matrix elements at zero and finite momentum, boosted in the $x$-direction.
We identified a set operators that contribute only at finite momentum and have particularly low excited-state contamination.
For the future, we are working on a direct 2-state analysis of the ratios to improve the quantitative analysis of the excited-state contamination.

\begin{acknowledgments}
  We thank the Budapest-Marseille-Wuppertal Collaboration for making
  their configurations available to us and Nesreen Hasan for calculating
  the correlation functions analysed here during the course of a different project.
  Calculations for this project
  were done using the Qlua software suite~\cite{Qlua}, and some of
  them made use of the QOPQDP adaptive multigrid
  solver~\cite{Babich:2010qb, QOPQDP}.  
  We gratefully acknowledge the computing time granted by the JARA Vergabegremium and provided on
  the JARA Partition part of the supercomputer JURECA~\cite{jureca} at
  Jülich Supercomputing Centre (JSC); computing time granted by the
  John von Neumann Institute for Computing (NIC) on the supercomputers
  JUQUEEN~\cite{juqueen}, JURECA, and JUWELS~\cite{juwels} at JSC; and
  computing time granted by the HLRS Steering Committee on Hazel Hen
  at the High Performance Computing Centre Stuttgart (HLRS).
  M.R.\@ was supported under the RWTH Exploratory Research Space (ERS) grant PF-JARA-SDS005 and MKW NRW under the funding code NW21-024-A.
  M.E.\@, J.N.\@ A.P.\@ are supported by the U.S. DOE Office of Science, Office of Nuclear Physics, through grant DE-FG02-96ER40965, DE-SC-0011090 and DE-SC0023116, respectively.
  S.M.\@ is supported by the U.S. Department of Energy, Office of Science, Office of High Energy Physics under Award Number DE-SC0009913.
  S.S.\@ is supported by the National Science Foundation under CAREER Award PHY-1847893"
\end{acknowledgments}
 
\appendix
\FloatBarrier

\bibliographystyle{apsrev4-1}

\end{document}